\documentclass[%
 reprint,
superscriptaddress,
 amsmath,amssymb,
 aps,
]{revtex4-2}

\usepackage{graphicx}
\usepackage{dcolumn}
\usepackage{bm}
\usepackage{hyperref}
\usepackage[all]{hypcap}
\hypersetup{
    colorlinks=true,
    linkcolor=red,
    citecolor=blue
    }
    
\usepackage[dvipsnames]{xcolor}
\usepackage{physics}
\usepackage{ulem}
\newcommand\soutc{\bgroup\markoverwith
{\textcolor{red}{\rule[.5ex]{2pt}{0.4pt}}}\ULon}

\newcommand{\ctb}{Ce$_3$TiBi$_5$}
\newcommand{\cts}{Ce$_3$TiSb$_5$}

\begin{document}


\title{The magnetic structure of \ctb\ and its relation to current-induced magnetization}


\author{Nicolas Gauthier}
\affiliation{Institut Quantique, Département de physique \& Regroupement Québécois sur les Matériaux de Pointe (RQMP), Université de Sherbrooke, Sherbrooke, Québec J1K 2R1, Canada}

\author{Romain Sibille}
\affiliation{Laboratory for Neutron Scattering and Imaging,
Paul Scherrer Institute, CH-5232 Villigen, Switzerland}

\author{Vladimir Pomjakushin}
\affiliation{Laboratory for Neutron Scattering and Imaging,
Paul Scherrer Institute, CH-5232 Villigen, Switzerland}

\author{\O ystein S. Fjellv\aa g}
\affiliation{Laboratory for Neutron Scattering and Imaging,
Paul Scherrer Institute, CH-5232 Villigen, Switzerland}
\affiliation{Department for Hydrogen Technology, Institute for Energy Technology, PO Box 40, NO-2027, Kjeller, Norway}

\author{James Fraser}
\affiliation{Département de Physique \& Regroupement Québécois sur les Matériaux de Pointe (RQMP), Université de Montréal, Montréal, Québec H3C 3J7, Canada}

\author{Mathieu Desmarais}
\affiliation{Département de Physique \& Regroupement Québécois sur les Matériaux de Pointe (RQMP), Université de Montréal, Montréal, Québec H3C 3J7, Canada}

\author{Andrea D. Bianchi}
\affiliation{Département de Physique \& Regroupement Québécois sur les Matériaux de Pointe (RQMP), Université de Montréal, Montréal, Québec H3C 3J7, Canada}

\author{Jeffrey A. Quilliam}
\affiliation{Institut Quantique, Département de physique \& Regroupement Québécois sur les Matériaux de Pointe (RQMP), Université de Sherbrooke, Sherbrooke, Québec J1K 2R1, Canada}

\date{\today}
\begin{abstract}

The control of magnetization using electric fields has been extensively studied in magnetoelectric multiferroic insulator materials. Changes in magnetization in bulk metals caused by electric currents have attracted less attention. The recently discovered metallic magnet \ctb\ has been reported to exhibit current-induced magnetization. Here we determined the magnetic structure of \ctb\ using neutron diffraction, aiming to understand the microscopic origin of this magnetoelectric phenomenon in a metal. We established that the antiferromagnetic order emerging below $T_N=5$~K is a cycloid order described by $P6_3/mcm.1'(0,0,g)00sss$ with small moment sizes of $0.50(2)~\mu_B$ and propagation vector ${\bf k}=(0,0,0.386)$. Surprisingly, the symmetry of this magnetic structure is inconsistent with the presence of current-induced magnetization and potential origins of this inconsistency with previous results are discussed. Additionally, our results suggest that moments order along their hard magnetic direction in \ctb, a phenomenon which has been observed in other Kondo systems. 
\end{abstract}
\maketitle

\textit{Introduction - } 
The possibility to induce magnetization with an electric field, or an electric polarization with a magnetic field is a well known effect in magnetoelectric multiferroic materials and of relevance to novel electronics~\cite{Spaldin2019}. These magnetoelectric effects can arise due to so-called toroidal orders. Toroidal orders are described by combined electric and magnetic orders and exists in various multiferroic insulators~\cite{Popov1999,Spaldin2008}. In contrast, few examples of toroidal orders in metals have been found. Interestingly, theoretical work has shown that, in metals without local inversion symmetry, toroidal order modifies the electronic band structure and leads to anisotropic Hall responses as well as current-induced magnetization~\cite{Hayami2014}. Experimentally, magnetization induced by an electric current in a metallic magnet has only been recently demonstrated for UNi$_4$B~\cite{Saito2018}. The reported vortex-like magnetic structure of UNi$_4$B generates a toroidal moment and has been considered to be a fundamental element in the theoretical description of this effect~\cite{Hayami2014}. However, magnetoelectric responses not expected by this theory were also observed and an alternative theory suggesting a more complex magnetic structure exists~\cite{Ishitobi2023}.

The \ctb\ compound is another interesting candidate to study magnetoelectric effects caused by toroidal order in a metal without local inversion symmetry.
This material, first synthesized only recently~\cite{Motoyama2018}, crystallizes in a hexagonal structure (space group $P6_3/mcm$) with Ce zigzag chains along the $c$-axis, as shown in Fig. 1b. It exhibits an antiferromagnetic order below $T_N=5.0$~K, as evidenced by resistivity, magnetic susceptibility and specific heat measurements~\cite{Motoyama2018}. Its large Sommerfeld coefficient of $\gamma = 210$~mJ/(K$^2$ Ce-mol) indicates heavy fermion behaviour. Its magnetic susceptibility exhibits a peculiar anisotropic temperature dependence~~\cite{Motoyama2018,Motoyama2023}. The susceptibility in the $ab$-plane is more than three times the $c$-axis susceptibility below 20~K, suggesting a preferred moment orientation in the $ab$-plane. Yet, the in-plane susceptibility does not reduce below $T_N$ as in a typical antiferromagnet but continues to increase with decreasing temperature. Instead, typical antiferromagnetic behavior is observed for the significantly smaller $c$-axis susceptibility, suggesting an ordering along the magnetic hard-axis~\cite{Kwasigroch2022}. 

Interestingly, it was observed that applying a current along the $a$-axis in the presence of the antiferromagnetic order induces a magnetization along the $c$-axis in \ctb~\cite{Shinozaki2020}. Although numerous isostructural analogues of \ctb\ are being investigated~\cite{Bollore1995,Moore2002a,Matin2017,Ritter2021,Murakami2017,Motoyama2020,Nakagawa2023}, only \ctb\ has been reported to host current-induced magnetization up to now. To explain the magnetoelectric effect in \ctb, a theoretical work suggested a potential magnetic structure, associated with a ferrotoroidal order~\cite{Hayami2022a}. Nevertheless, its experimental determination is required to obtain a convincing understanding of this phenomenon. In this Letter, we report neutron diffraction measurements of \ctb\ and determine its magnetic structure, establishing that it is different from the theoretical proposal. Our results resolve ambiguities associated with its magnetic susceptibility but raise doubts about the intrinsic nature of the observed current-induced magnetization.


\textit{Methods - } 
Single crystals of \ctb\ were grown using the Bi self-flux method~\cite{Motoyama2018}. 
Starting material of cerium (99.9\%), titanium (99.99\%) and bismuth (99.999\%) in a ratio 3:1:20 were placed in an alumina crucible. Growth batch A was obtained by sealing the crucible in a quartz ampoule with argon. The ampoule was heated to 1000$^\circ$C at 50$^\circ$C/h, kept at this temperature for 11h and cooled down to 500$^\circ$C at a rate of 2$^\circ$C/h. The Bi excess was removed by centrifugation immediately after removing the ampoule from the oven. The obtained needle-shaped crystals are air-sensitive: in air they decompose in a few days and burn when broken or crushed. They were kept and handled in a controlled atmosphere in a glovebox, and transported to experiments in sealed containers. The \ctb\ phase was confirmed by x-ray powder diffraction using a domed sample holder with crystals powderized in the glovebox. Several large $\sim 50$~mg single crystals were obtained in this batch and one with dimensions $\sim5 \times 1\times 1$~mm$^3$ was used for the single crystal neutron diffraction experiment. Growth batch B was obtained by sealing the alumina crucible in a quartz ampoule under vacuum. In that case, the ampoule was kept at 1000$^\circ$C for 15h and cooled down at a rate of 1$^\circ$C/h, with all the other conditions identical to batch A. The resulting crystals were slightly smaller than in batch A. Needle-shaped crystals were isolated and powderized in a glovebox. The obtained 3.4~g powder was used for the powder neutron diffraction experiments.

Powder neutron diffraction measurements were performed using the HRPT and DMC instruments at SINQ, Paul Scherrer Institut. The 3.4~g powder was loaded in a vanadium can and sealed in a helium atmosphere. The sample was loaded in an orange cryostat and data were collected using neutron wavelengths $\lambda=2.45$~\AA\ and 3.81~\AA\ for HRPT and DMC instruments, respectively. Single crystal diffraction measurements were performed on the ZEBRA diffractometer at SINQ. The 50~mg sample was mounted in a Joule-Thomson cryostat in a 4-circle geometry and measured with $\lambda=1.383$~\AA\ neutrons.
Data were analyzed with the FULLPROF program~\cite{Fullprof} and figures of crystallographic and magnetic structures were generated using the VESTA software~\cite{VESTA}.

\begin{figure}[!htb]
\begin{center}
\includegraphics[scale=0.55]{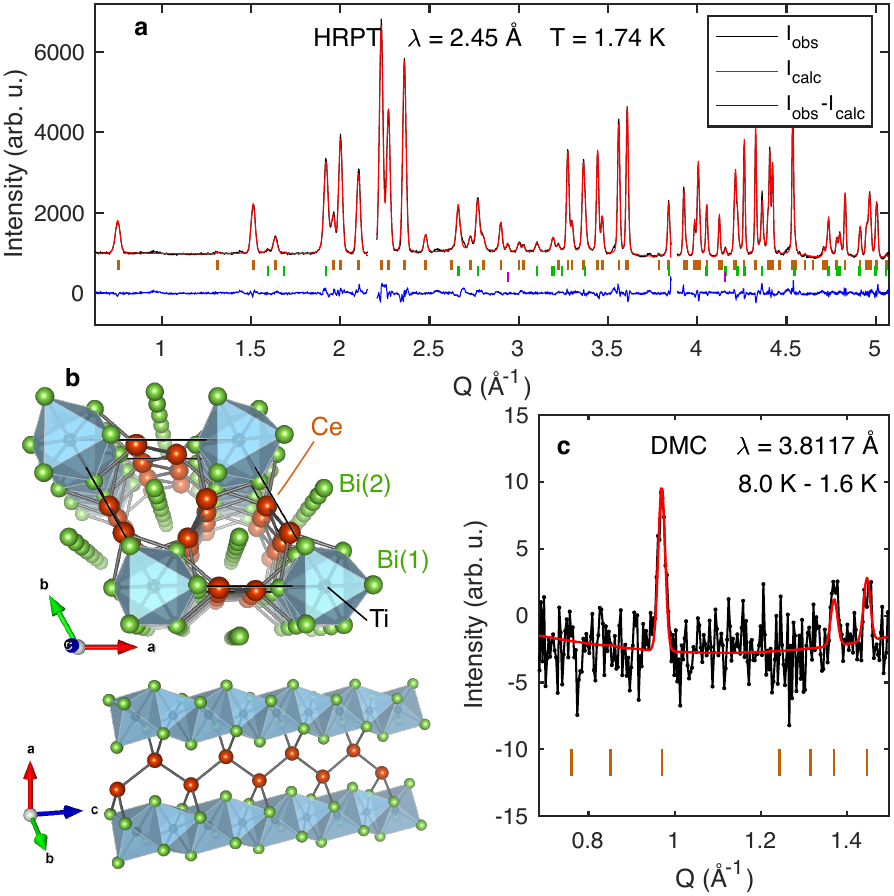}
\end{center}
\caption{({\bf a}) Powder neutron diffraction pattern of \ctb\ measured with the HRPT diffractometer. Dark orange and green ticks indicate peak positions of the \ctb\ phase and a Bi impurity phase, respectively. Purple ticks indicate vanadium peaks from the sample container. Regions at $\sim${2.2~\AA$^{-1}$} and $\sim${3.9~\AA$^{-1}$} are associated with dominant peaks of CeTi$_3$Bi$_4$ impurities and were excluded.
({\bf b}) Crystallographic structure of \ctb, showing the Ce zigzag chains along the $c$-axis.
({\bf c}) Difference powder pattern between 8.0~K and 1.6~K measured with the DMC diffractometer. Three magnetic peaks are observed and are well described by the $P6_3/mcm.1'(0,0,g)00sss$ magnetic structure.}
\label{fig1}
\end{figure}

\textit{Experimental results - } 
High resolution powder neutron diffraction patterns were measured using the HRPT diffractometer at  $T=1.74$~K and 8~K, i.e. below and above $T_N$. The pattern and associated refinement for $T=1.74$~K is presented in Fig.~\ref{fig1}a. The pattern is dominated by the \ctb\ phase but a large weight fraction of Bi impurity phase ($\sim18\%$) is present, likely originating from excess Bi flux. The refinement of the  \ctb\ confirms the previously established $P6_3/mcm$ structure~\cite{Motoyama2018}, illustrated in Fig.~\ref{fig1}b. The refined crystallographic parameters for both temperatures are reported in Table~\ref{tab1}. There is no evidence for any structural distortion. Except for a small contraction along the $c$-axis, all refined parameters are temperature independent within the measurement uncertainties. No new peaks associated with magnetic order were observed for $T<T_N$ in this measurement due to weak magnetic scattering at short neutron wavelength.

\begin{table}
\begin{tabular}{l || c c} 
 \hline
T (K) & 1.74 & 8.0 \\
 \hline 
 $a$ (\AA)  & \ \ 9.58146(4) \ \ & \ \ 9.58140(5) \ \ \\
$c$ (\AA)  &  6.40091(4)  & 6.40106(4) \\
Vol. (\AA$^3$)  & 508.904(4) &  508.909(4) \\
 \hline 
$x_{\text{Ce}}$   & 0.6179(2) &  0.6177(2) \\
$x_{\text{Bi(1)}}$  \ \ & 0.25613(9)&  0.2562(1) \\
 \hline 
$R_\text{wp}$ & 7.11 &   8.29 \\
$R_\text{exp}$ & 2.91 &  5.47 \\
$\chi^2$ & 5.96 &  2.30 \\
$R_\text{Bragg}$ & 2.33 &  2.20 \\
\hline
\end{tabular}
\caption{Crystallographic parameters obtained from refinement of HRPT diffraction patterns. Atomic positions are 
Ce at $6g\ (x,0,1/4)$, Ti at $2b\ (0,0,0)$, Bi(1) at $6g\ (x,0,1/4)$ and Bi(2) at $4d\ (1/3,2/3,0)$. Data obtained 1.74~K have 3.5 times more statistics than the 8~K data, justifying the difference in $\chi^2$ between datasets.}
\label{tab1}
\end{table}

\begin{figure}[!htb]
\begin{center}
\includegraphics[scale=0.55]{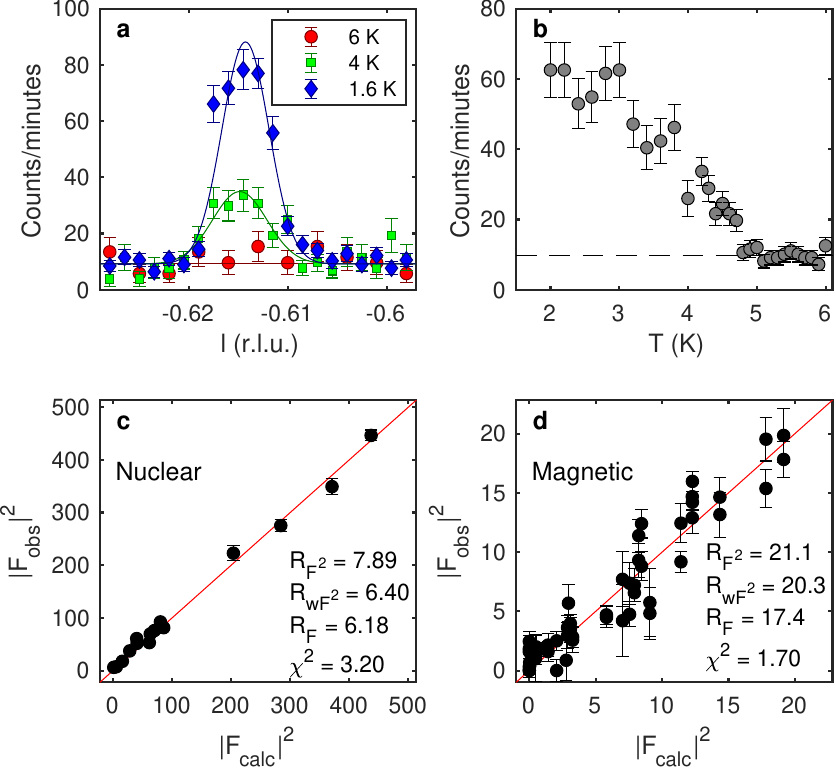}
\end{center}
\caption{({\bf a}) Temperature dependence of the magnetic Bragg peak ${\bf Q} =(-1,1,l)$, showing that the incommensurate value does not change with temperature. Data were fitted with a Gaussian function. 
({\bf b}) Temperature dependence of the scattering intensity at the Bragg peak position ${\bf Q} =(-1,1,-0.614)$ showing the emergence of the magnetic order parameter at $T_N=5$~K. 
({\bf c}) Result of the nuclear structure refinement at $T=1.6$~K, represented by the comparison of $| F_\text{obs}|^2$ and $| F_\text{calc}|^2$, the observed and calculated squared structure factors, respectively. 
({\bf d}) Result of the $P6_3/mcm.1'(0,0,g)00sss$ magnetic structure refinement in the same conditions. 
}
\label{fig2}
\end{figure}

Magnetic peaks were revealed by measurements on the DMC diffractometer, optimized for magnetic structure determination. A difference pattern of 1.6~K and 8~K data, presented in Fig.~\ref{fig1}c, exhibits a clear peak at $Q=0.97~$\AA$^{-1}$ as well as two weaker peaks around 1.4~\AA$^{-1}$. These peaks are insufficient to establish unambiguously the propagation vector associated with the antiferromagnetic order. The propagation vector ${\bf k}=(0,0,\delta)$ with $\delta=0.386$ was determined by single crystal diffraction with the ZEBRA diffractometer. Fig.~\ref{fig2}a shows the temperature dependence of the magnetic Bragg peak ${\bf Q}=(-1,1,-0.614)$. The peak position is unaffected by temperature, indicating that the incommensurate parameter $\delta$ is temperature independent. The peak intensity decreases when increasing temperature from 1.6~K to 4~K and is absent at $T= 6$~K$>T_N$.  The intensity at the top of the magnetic Bragg peak decreases with increasing temperature and reaches the background intensity at $T_N$ (Fig.~\ref{fig2}b), confirming that it arises from the magnetic order.

Nuclear and magnetic Bragg peaks were collected at $T=1.6~K$ to perform crystallographic and magnetic structure refinements using the FULLPROF program~\cite{Fullprof}. The scaling factor obtained from the nuclear dataset refinement, shown in Fig.~\ref{fig2}c, was used to determine absolute moment sizes. Magnetic peaks were collected for both $+{\bf k}$ and $-{\bf k}$ and refined jointly. Possible magnetic structures were determined from a symmetry analysis using the ISODISTORT tool~\cite{StokesISOTROPY,Campbell2006}. The propagation vector $\bf k$ corresponds to the $\Delta$ point of the Brillouin zone and leads to six potential magnetic irreducible representations labeled $m\Delta_i$ with $i=1$ to 6. Considering the incommensurate nature of $\bf k$, we describe the magnetic structure using the magnetic superspace group (MSG) formalism~\cite{Perez-Mato2012}. 
Superspace groups go beyond the usual three-dimensional space groups by assuming one or more additional dimensions to treat incommensurate structures. Magnetic space groups consider the time-reversal symmetry operator to account for magnetism. MSGs combine both concepts.
The DMC difference pattern is well accounted for by considering MSGs generated by either $m\Delta_2$ or $m\Delta_5$. The acceptable $m\Delta_5$-MSGs all contain the $m\Delta_2$ distortion modes but have a lower symmetry than the $m\Delta_2$-MSGs. Therefore, the $m\Delta_5$-MSGs are ignored to keep only the highest symmetry solutions generated by $m\Delta_2$. The candidate MSGs are $P6_3/mcm.1'(0,0,g)00sss$ and $P6_3cm.1'(0,0,g)0sss$, with the main difference being the loss of inversion symmetry in the latter. 

\begin{figure}[!htb]
\begin{center}
\includegraphics[scale=0.55]{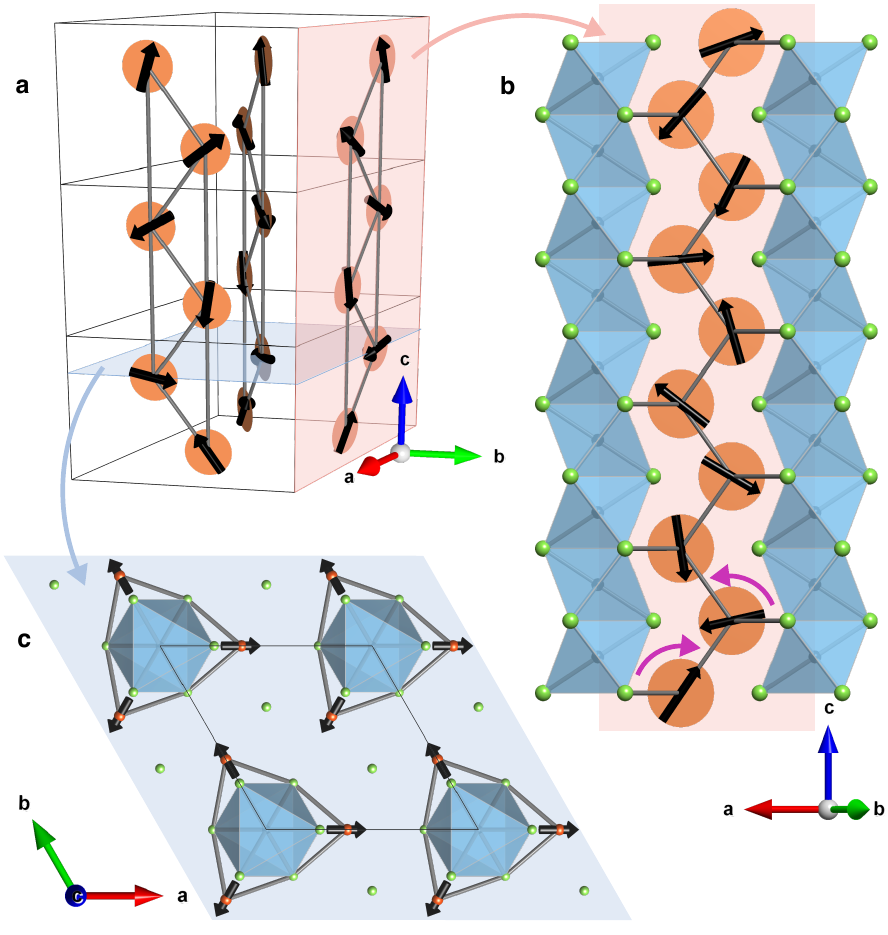}
\end{center}
\caption{({\bf a}) Magnetic structure of \ctb. For each zigzag chain, the moments are confined in a plane, represented by the orange circle at each Ce site. The $ac$ plane indicated in red and the $ab$ plane indicated in blue are presented in ({\bf b}) and ({\bf c}), respectively. The ordering on the zigzag chain is explicitly presented in ({\bf b}). The two sides of the zigzag chain form a cycloid order but with opposite rotation directions. The moments on left side of the zigzag chain rotate clockwise from one site to the next along the $c$-axis, as illustrated by the purple arrow. In contrast, the moments on the right side rotate counterclockwise.  The $ab$ plane presented in ({\bf c}) illustrates the ordering of the in-plane magnetic components between the chains.  
}
\label{fig3}
\end{figure}

Refinement of the ZEBRA magnetic datasets also indicates that the two $m\Delta_2$-generated MSGs are the best solutions with 
agreement factors $R_F=17.4$ and 21.0
for $P6_3/mcm.1'(0,0,g)00sss$ and $P6_3cm.1'(0,0,g)0sss$, respectively. 
{Comparable agreement factors can only be obtained with $m\Delta_5$-generated MSGs, which are disregarded in the following due to their lower symmetry.} 
Considering that both {$m\Delta_2$} solutions provide comparable agreement with the data, the first, higher symmetry $P6_3/mcm.1'(0,0,g)00sss$ structure is assumed to be the correct description of \ctb\ antiferromagnetic order. The refinement result for this MSG is shown in Fig.~\ref{fig2}d and the obtained magnetic structure is presented in Fig.~\ref{fig3}. It corresponds to a cycloid order propagating along the $c$-axis. For each zigzag chain, the moments lie in the plane containing that zigzag chain. The incommensurate value $\delta$ leads to a moment rotation of 139$^\circ$ per unit cell along the $c$-axis. It is noteworthy that the moments on the two legs of the zigzag chain rotate in opposite directions{, and models with moments rotating in the same direction cannot describe our data satisfactorily}. 
For the zigzag chain illustrated in Fig.~\ref{fig3}b, the left side of the chain rotates clockwise and the right side of the chain rotates counterclockwise, as indicated by the purple arrows. 
The relative phase between these opposite rotations is symmetry-enforced. 
This leads to moment components along the $c$-axis that are predominantly aligned ferromagnetically for nearest neighbor bonds, while the transverse moment components approach an antiferromagnetic alignment. {Such alignment suggests that anisotropic interactions are responsible for the opposite rotations and their relative phase. We note that counterrotating cycloids were also reported in iridates with anisotropic Kitaev interactions~\cite{Biffin2014}. } 

The evolution of the magnetic moments ${\bf M}_{i}$ along the propagation axis can be expressed by 
\begin{equation}
{\bf M}_{i}=  {\bf M}_{i}^{\sin} \sin{\left(2 \pi \delta Z_i \right)} +{\bf M}_{i}^{\cos} \cos{\left(2 \pi \delta Z_i \right)} 
\end{equation} 
where $Z_i$ is the position along $c$ of site $i$ in lattice units. Moments ${\bf M}_{i}$ within one unit cell are related to each other by the symmetry operators of the MSGs. 
For the $P6_3/mcm.1'(0,0,g)00sss$ group, ${\bf M}_{i}^{\cos}$ and ${\bf M}_{i}^{\sin}$ are constrained to be orthogonal and lie in the $xz$ plane. Here, $x$ is along the local 2-fold axis while $z$ is along the $c$-axis.  Specifically, the magnetic components are given by ${\bf M}_{i}^{\sin} = (M^{\sin}_x,0,0)$ and ${\bf M}_{i}^{\cos} = (0,0,M^{\cos}_z)$. This describes a conventional cycloid order when $M^{\sin}_x$ and $M^{\cos}_z$ are accidentally equal. If $M^{\sin}_x\neq M^{\cos}_z$, it corresponds to a cycloid order with an ellipsoidal enveloppe having principal axes along $x$ and $z$. Our single-crystal data refinement presented in Fig.~\ref{fig2}d suggests a conventional cycloid order with $M^{\sin}_x=-0.50(2)\ \mu_B$ and $M^{\cos}_z=-0.50(2)\ \mu_B$. 
The refinement of the DMC difference pattern with the same MSG (red line on Fig.~\ref{fig1}c) indicates slightly smaller moment sizes of $M^{\sin}_x= -0.40(2)\ \mu_B$ and $M^{\cos}_z=0.37(3)\ \mu_B$ with an incommensurate value $\delta=0.389(2)$. The goodness of fit and $R$-factors for this refinement are $R_\text{wp}=75.0$, $R_\text{exp}=78.9$, $\chi^2=0.91$, $R_\text{Bragg}=21.4$.

For the lower symmetry $P6_3cm.1'(0,0,g)0sss$ group, which breaks inversion symmetry, ${\bf M}_{i}^{\cos}$ and ${\bf M}_{i}^{\sin}$ are still constrained to lie in the $xz$ plane but the orthogonality restriction is removed. Specifically, the magnetic components are given by ${\bf M}_{i}^{\sin} = (M^{\sin}_x,0,M^{\sin}_z)$ and ${\bf M}_{i}^{\cos} = (M^{\cos}_x,0,M^{\cos}_z)$. The added degree of freedom of this MSG compared to $P6_3/mcm.1'(0,0,g)00sss$ allows the generation of a cycloid order with an ellipsoidal enveloppe with principal axes away from $x$ and $z$, or of a non-cycloid amplitude-modulated order. For this MSG, the single-crystal data refinement   leads to magnetic components describing a structure that approximates the cycloid order obtained for the higher symmetry $P6_3/mcm.1'(0,0,g)00sss$ group. $\chi^2$ values were evaluated for various angles $\theta$ between ${\bf M}_{i}^{\sin} $ and $ {\bf M}_{i}^{\cos}$ while constraining equal amplitudes $| {\bf M}_{i}^{\sin} | = | {\bf M}_{i}^{\cos} |$. For $\theta=90^\circ$, the structure is identical to the higher symmetry $P6_3/mcm.1'(0,0,g)00sss$ solution. We find that $\chi^2$ values are almost unchanged for $60^\circ< \theta < 120^\circ$ and increase further away from orthogonality. Therefore, a small distortion of the magnetic structure breaking inversion symmetry would not be in disagreement with our results. 
{We emphasize that the dominant character of the structure remains a cycloid order with moments lying in the plane of the zigzag chain. The same dominant cycloid character is obtained for all the lower-symmetry $m\Delta_5$-generated MSGs that provided comparable agreement factors to the $m\Delta_2$ solutions.}

 \textit{Discussion - }    
{ Surprisingly, the observation of a single-$k$ incommensurate magnetic order in \ctb\ indicates directly that the current-induced magnetization effect is not allowed by symmetry, in clear contradiction with reported experimental results~\cite{Shinozaki2020}. The presence of a ferrotoroidal order is requisite to the current-induced magnetization effect according to the theoretical model~\cite{Hayami2022a}, and such orders are only allowed when both inversion and time reversal symmetries are broken by the magnetic point group. However, the magnetic point group of single-$k$ incommensurate phases always includes time reversal symmetry and consequently cannot have a ferrotoroidal order~\cite{Perez-Mato2012}. In other words, ferrotoroidal order is forbidden in \ctb\ simply by the observation of a single incommensurate propagation vector, regardless of the presence of inversion symmetry, the details of the magnetic structure and its exact MSG. It is therefore hard to reconcile the reported magnetoelectric effect~\cite{Shinozaki2020} with the observed single-$k$ incommensurate magnetic structure. There is no evidence for sample or growth batch dependent properties, either in previous results or in our neutron scattering results, that could explain this inconsistency. Specifically in this work, the results from powder and single crystal measurements performed on two different growth batches are in good agreement.} 
 
{Ferrotoroidal order would be allowed if an undetected second $k$-vector exists and forms a multi-$k$ magnetic structure breaking both inversion and time-reversal symmetry. However, we found no evidence for a second $k$-vector. All peaks in the powder diffraction pattern are accounted for. Furthermore, using single crystal diffraction we measured all points within the irreducible Brillouin zone at $Q=0.97$~\AA$^{-1}$, where the strongest magnetic scattering is observed in the powder pattern, and magnetic scattering was only observed for ${\bf k}=(0,0,\delta)$. Based on this and the good agreement between the magnetic structures determined for powder and single crystal samples, we can exclude any significant contribution from an additional $k$-vector at this $Q$-value. Yet, we cannot generally exclude the existence of an additional $k$-vector with weak scattering intensity at any $Q$-value.}
 
{While a second $k$-vector cannot be fully excluded, our results bring into question} 
 the intrinsic nature of the current-induced magnetization effect observed in the challenging experiment reported in Ref.~\cite{Shinozaki2020}. In the analysis of this work, the temperature independent part of the current-induced magnetization was attributed to an extrinsic effect, while the temperature dependent part was associated with the sample's response. 
However, the extrinsic induced fields depend, in part, on the path of the current inside the sample. With an imperfect contact geometry, this path is affected by the anisotropy of the resistivity and could be modified with temperature if the anisotropy is temperature dependent. In \ctb, the resistivity anisotropy ratio is $\rho_c/\rho_{ab}\approx 0.2$ between 5 and 10~K while it suddenly drops below $T_N$, reaching $\rho_c/\rho_{ab}\approx 0.1$ at 2~K~\cite{Motoyama2018}. In brief, significant changes of the resistivity anisotropy occur simultaneously with the magnetic order and might generate temperature dependent extrinsic fields that could be erroneously interpreted as an intrinsic property of the material. Current-induced measurements on samples with different shapes ought to be carried out to investigate this possibility.

While our results indicate that \ctb\ does not exhibit a ferrotoroidal order and that the current-induced magnetization could be an experimental artefact, its magnetic properties appear unconventional.
Interestingly, the established magnetic structure of \ctb, together with its peculiar magnetic susceptibility, suggests that \ctb\ might be another example of an antiferromagnetic Kondo system ordering along the magnetic hard direction~\cite{Hafner2019}. Generally speaking, ordering of moments is expected along the easy axis for isotropic interactions, as the energy gain is proportional to the square of the moment size. However, it was shown theoretically that particle-hole fluctuations in metallic ferromagnets can generate magnetic orders with moments along their hard axes, analogous to the stabilization of an inverted pendulum with vibrations~\cite{Kruger2014}. This counterintuitive phenomenon is in fact common in Kondo ferromagnets, while only a few Kondo antiferromagnets have been shown to fall into this category~\cite{Hafner2019,Kwasigroch2022}. In \ctb, 
the peculiar magnetic susceptibility can be qualitatively explained by assuming that for each zigzag chain the local hard plane corresponds to the plane of the zigzag chain, and the local easy axis is perpendicular to it.
The $c$-axis susceptibility, probing one component of the hard plane, is weak and clearly exhibits a drop at the AFM transition as this component orders. The in-plane susceptibility probes simultaneously the other hard-plane component and the easy-axis one, but is dominated by the easy-axis component which has a larger susceptibility. As this easy-axis component does not order, the susceptibility continues to grow below $T_N$ while the kink is caused by the weaker hard-plane component. 

Another indication of hard-direction ordering is given by the magnetization resulting from fields applied in the $ab$-plane. Indeed, the magnetization starts to saturate at $\sim 1$~$\mu_B$/Ce above $H=7$~T~\cite{Motoyama2023}, a value that is twice the fully ordered moment in zero field determined here. The polarization of the moments along the easy axis by the field provides a simple explanation for this large magnetization. It cannot be explained by field-induced mixing of higher energy CEF levels, as they are separated by tens of K from the ground state doublet based on specific heat measurements~\cite{Motoyama2023}. In contrast to \ctb, the Sb-analogue \cts\ exhibits a more conventional magnetic susceptibility, with its AFM transition observed in the largest susceptibility component~\cite{Matin2017}. Powder neutron diffraction measurements performed on \cts\ at $T=1.5$~K in the AFM phase revealed large moment sizes of $\sim2 \mu_B$ and a dominant moment direction transverse to the plane of the zigzag chain~\cite{Ritter2021}. This suggests that \cts\ and \ctb\ have comparable anisotropies but that \cts\ orders along the easy axis while Kondo fluctuations drive an ordering in the hard plane for \ctb.

In summary, we established the magnetic structure of \ctb\ to be a cycloid order with moments lying in the plane of the zigzag chain. This structure does not allow a ferrotoroidal order and calls into question the intrinsic nature of the reported current-induced magnetization effect. The determined magnetic structure, together with the previously measured magnetic susceptibility, suggests that this material exhibits an order along its magnetic hard direction, comparable to what has been observed in other Kondo systems. An accurate determination of the CEF parameters, from techniques such as neutron spectroscopy, is required to establish the single-ion moment anisotropy that could support the hard-plane ordering scenario. 
Furthermore, the role of Kondo fluctuations in this peculiar phenomenon could be evidenced by chemical substitution, 
e.g. by replacing itinerant Ce moments with localized Nd moments~\cite{Mazzone2019a}.

\section*{}
This work is based on experiments performed at the Swiss spallation neutron source SINQ, Paul Scherrer Institute, Villigen, Switzerland. N.G. and J.A.Q. acknowledge the support of the Canada First Research Excellence Fund (CFREF). \O .S.F. acknowledges funding from Research Council of Norway (NFR) through project 325345.

%


\begin{thebibliography}{27}%
\makeatletter
\providecommand \@ifxundefined [1]{%
 \@ifx{#1\undefined}
}%
\providecommand \@ifnum [1]{%
 \ifnum #1\expandafter \@firstoftwo
 \else \expandafter \@secondoftwo
 \fi
}%
\providecommand \@ifx [1]{%
 \ifx #1\expandafter \@firstoftwo
 \else \expandafter \@secondoftwo
 \fi
}%
\providecommand \natexlab [1]{#1}%
\providecommand \enquote  [1]{``#1''}%
\providecommand \bibnamefont  [1]{#1}%
\providecommand \bibfnamefont [1]{#1}%
\providecommand \citenamefont [1]{#1}%
\providecommand \href@noop [0]{\@secondoftwo}%
\providecommand \href [0]{\begingroup \@sanitize@url \@href}%
\providecommand \@href[1]{\@@startlink{#1}\@@href}%
\providecommand \@@href[1]{\endgroup#1\@@endlink}%
\providecommand \@sanitize@url [0]{\catcode `\\12\catcode `\$12\catcode
  `\&12\catcode `\#12\catcode `\^12\catcode `\_12\catcode `\%12\relax}%
\providecommand \@@startlink[1]{}%
\providecommand \@@endlink[0]{}%
\providecommand \url  [0]{\begingroup\@sanitize@url \@url }%
\providecommand \@url [1]{\endgroup\@href {#1}{\urlprefix }}%
\providecommand \urlprefix  [0]{URL }%
\providecommand \Eprint [0]{\href }%
\providecommand \doibase [0]{https://doi.org/}%
\providecommand \selectlanguage [0]{\@gobble}%
\providecommand \bibinfo  [0]{\@secondoftwo}%
\providecommand \bibfield  [0]{\@secondoftwo}%
\providecommand \translation [1]{[#1]}%
\providecommand \BibitemOpen [0]{}%
\providecommand \bibitemStop [0]{}%
\providecommand \bibitemNoStop [0]{.\EOS\space}%
\providecommand \EOS [0]{\spacefactor3000\relax}%
\providecommand \BibitemShut  [1]{\csname bibitem#1\endcsname}%
\let\auto@bib@innerbib\@empty
\bibitem [{\citenamefont {Spaldin}\ and\ \citenamefont
  {Ramesh}(2019)}]{Spaldin2019}%
  \BibitemOpen
  \bibfield  {author} {\bibinfo {author} {\bibfnamefont {N.~A.}\ \bibnamefont
  {Spaldin}}\ and\ \bibinfo {author} {\bibfnamefont {R.}~\bibnamefont
  {Ramesh}},\ }\bibfield  {title} {\bibinfo {title} {{Advances in
  magnetoelectric multiferroics}},\ }\href
  {https://doi.org/10.1038/s41563-018-0275-2} {\bibfield  {journal} {\bibinfo
  {journal} {Nature Materials}\ }\textbf {\bibinfo {volume} {18}},\ \bibinfo
  {pages} {203} (\bibinfo {year} {2019})}\BibitemShut {NoStop}%
\bibitem [{\citenamefont {Popov}\ \emph {et~al.}(1999)\citenamefont {Popov},
  \citenamefont {Kadomtseva}, \citenamefont {Belov}, \citenamefont {Vorob'ev},\
  and\ \citenamefont {Zvezdin}}]{Popov1999}%
  \BibitemOpen
  \bibfield  {author} {\bibinfo {author} {\bibfnamefont {Y.~F.}\ \bibnamefont
  {Popov}}, \bibinfo {author} {\bibfnamefont {A.~M.}\ \bibnamefont
  {Kadomtseva}}, \bibinfo {author} {\bibfnamefont {D.~V.}\ \bibnamefont
  {Belov}}, \bibinfo {author} {\bibfnamefont {G.~P.}\ \bibnamefont
  {Vorob'ev}},\ and\ \bibinfo {author} {\bibfnamefont {A.~K.}\ \bibnamefont
  {Zvezdin}},\ }\bibfield  {title} {\bibinfo {title} {{Magnetic-field-induced
  toroidal moment in the magnetoelectric Cr$_2$O$_3$}},\ }\href
  {https://doi.org/10.1134/1.568032} {\bibfield  {journal} {\bibinfo  {journal}
  {Journal of Experimental and Theoretical Physics Letters}\ }\textbf {\bibinfo
  {volume} {69}},\ \bibinfo {pages} {330} (\bibinfo {year} {1999})}\BibitemShut
  {NoStop}%
\bibitem [{\citenamefont {Spaldin}\ \emph {et~al.}(2008)\citenamefont
  {Spaldin}, \citenamefont {Fiebig},\ and\ \citenamefont
  {Mostovoy}}]{Spaldin2008}%
  \BibitemOpen
  \bibfield  {author} {\bibinfo {author} {\bibfnamefont {N.~A.}\ \bibnamefont
  {Spaldin}}, \bibinfo {author} {\bibfnamefont {M.}~\bibnamefont {Fiebig}},\
  and\ \bibinfo {author} {\bibfnamefont {M.}~\bibnamefont {Mostovoy}},\
  }\bibfield  {title} {\bibinfo {title} {{The toroidal moment in
  condensed-matter physics and its relation to the magnetoelectric effect}},\
  }\href {https://doi.org/10.1088/0953-8984/20/43/434203} {\bibfield  {journal}
  {\bibinfo  {journal} {Journal of Physics: Condensed Matter}\ }\textbf
  {\bibinfo {volume} {20}},\ \bibinfo {pages} {434203} (\bibinfo {year}
  {2008})}\BibitemShut {NoStop}%
\bibitem [{\citenamefont {Hayami}\ \emph {et~al.}(2014)\citenamefont {Hayami},
  \citenamefont {Kusunose},\ and\ \citenamefont {Motome}}]{Hayami2014}%
  \BibitemOpen
  \bibfield  {author} {\bibinfo {author} {\bibfnamefont {S.}~\bibnamefont
  {Hayami}}, \bibinfo {author} {\bibfnamefont {H.}~\bibnamefont {Kusunose}},\
  and\ \bibinfo {author} {\bibfnamefont {Y.}~\bibnamefont {Motome}},\
  }\bibfield  {title} {\bibinfo {title} {{Toroidal order in metals without
  local inversion symmetry}},\ }\href
  {https://doi.org/10.1103/PhysRevB.90.024432} {\bibfield  {journal} {\bibinfo
  {journal} {Physical Review B}\ }\textbf {\bibinfo {volume} {90}},\ \bibinfo
  {pages} {024432} (\bibinfo {year} {2014})}\BibitemShut {NoStop}%
\bibitem [{\citenamefont {Saito}\ \emph {et~al.}(2018)\citenamefont {Saito},
  \citenamefont {Uenishi}, \citenamefont {Miura}, \citenamefont {Tabata},
  \citenamefont {Hidaka}, \citenamefont {Yanagisawa},\ and\ \citenamefont
  {Amitsuka}}]{Saito2018}%
  \BibitemOpen
  \bibfield  {author} {\bibinfo {author} {\bibfnamefont {H.}~\bibnamefont
  {Saito}}, \bibinfo {author} {\bibfnamefont {K.}~\bibnamefont {Uenishi}},
  \bibinfo {author} {\bibfnamefont {N.}~\bibnamefont {Miura}}, \bibinfo
  {author} {\bibfnamefont {C.}~\bibnamefont {Tabata}}, \bibinfo {author}
  {\bibfnamefont {H.}~\bibnamefont {Hidaka}}, \bibinfo {author} {\bibfnamefont
  {T.}~\bibnamefont {Yanagisawa}},\ and\ \bibinfo {author} {\bibfnamefont
  {H.}~\bibnamefont {Amitsuka}},\ }\bibfield  {title} {\bibinfo {title}
  {{Evidence of a New Current-Induced Magnetoelectric Effect in a Toroidal
  Magnetic Ordered State of UNi$_4$B}},\ }\href
  {https://doi.org/10.7566/JPSJ.87.033702} {\bibfield  {journal} {\bibinfo
  {journal} {Journal of the Physical Society of Japan}\ }\textbf {\bibinfo
  {volume} {87}},\ \bibinfo {pages} {033702} (\bibinfo {year}
  {2018})}\BibitemShut {NoStop}%
\bibitem [{\citenamefont {Ishitobi}\ and\ \citenamefont
  {Hattori}(2023)}]{Ishitobi2023}%
  \BibitemOpen
  \bibfield  {author} {\bibinfo {author} {\bibfnamefont {T.}~\bibnamefont
  {Ishitobi}}\ and\ \bibinfo {author} {\bibfnamefont {K.}~\bibnamefont
  {Hattori}},\ }\bibfield  {title} {\bibinfo {title} {{Triple-Q partial
  magnetic orders induced by quadrupolar interactions: Triforce order scenario
  for UNi$_4$B}},\ }\href {https://doi.org/10.1103/PhysRevB.107.104413}
  {\bibfield  {journal} {\bibinfo  {journal} {Physical Review B}\ }\textbf
  {\bibinfo {volume} {107}},\ \bibinfo {pages} {104413} (\bibinfo {year}
  {2023})}\BibitemShut {NoStop}%
\bibitem [{\citenamefont {Motoyama}\ \emph {et~al.}(2018)\citenamefont
  {Motoyama}, \citenamefont {Sezaki}, \citenamefont {Gouchi}, \citenamefont
  {Miyoshi}, \citenamefont {Nishigori}, \citenamefont {Mutou}, \citenamefont
  {Fujiwara},\ and\ \citenamefont {Uwatoko}}]{Motoyama2018}%
  \BibitemOpen
  \bibfield  {author} {\bibinfo {author} {\bibfnamefont {G.}~\bibnamefont
  {Motoyama}}, \bibinfo {author} {\bibfnamefont {M.}~\bibnamefont {Sezaki}},
  \bibinfo {author} {\bibfnamefont {J.}~\bibnamefont {Gouchi}}, \bibinfo
  {author} {\bibfnamefont {K.}~\bibnamefont {Miyoshi}}, \bibinfo {author}
  {\bibfnamefont {S.}~\bibnamefont {Nishigori}}, \bibinfo {author}
  {\bibfnamefont {T.}~\bibnamefont {Mutou}}, \bibinfo {author} {\bibfnamefont
  {K.}~\bibnamefont {Fujiwara}},\ and\ \bibinfo {author} {\bibfnamefont
  {Y.}~\bibnamefont {Uwatoko}},\ }\bibfield  {title} {\bibinfo {title}
  {{Magnetic properties of new antiferromagnetic heavy-fermion compounds,
  Ce$_3$TiBi$_5$ and CeTi$_3$Bi$_4$}},\ }\href
  {https://doi.org/10.1016/j.physb.2017.10.005} {\bibfield  {journal} {\bibinfo
   {journal} {Physica B: Condensed Matter}\ }\textbf {\bibinfo {volume}
  {536}},\ \bibinfo {pages} {142} (\bibinfo {year} {2018})}\BibitemShut
  {NoStop}%
\bibitem [{\citenamefont {Motoyama}\ \emph {et~al.}(2023)\citenamefont
  {Motoyama}, \citenamefont {Shinozaki}, \citenamefont {Nishigori},
  \citenamefont {Yamaguchi}, \citenamefont {Aso}, \citenamefont {Mutou},
  \citenamefont {Manago}, \citenamefont {Fujiwara}, \citenamefont {Sumiyama},\
  and\ \citenamefont {Uwatoko}}]{Motoyama2023}%
  \BibitemOpen
  \bibfield  {author} {\bibinfo {author} {\bibfnamefont {G.}~\bibnamefont
  {Motoyama}}, \bibinfo {author} {\bibfnamefont {M.}~\bibnamefont {Shinozaki}},
  \bibinfo {author} {\bibfnamefont {S.}~\bibnamefont {Nishigori}}, \bibinfo
  {author} {\bibfnamefont {A.}~\bibnamefont {Yamaguchi}}, \bibinfo {author}
  {\bibfnamefont {N.}~\bibnamefont {Aso}}, \bibinfo {author} {\bibfnamefont
  {T.}~\bibnamefont {Mutou}}, \bibinfo {author} {\bibfnamefont
  {M.}~\bibnamefont {Manago}}, \bibinfo {author} {\bibfnamefont
  {K.}~\bibnamefont {Fujiwara}}, \bibinfo {author} {\bibfnamefont
  {A.}~\bibnamefont {Sumiyama}},\ and\ \bibinfo {author} {\bibfnamefont
  {Y.}~\bibnamefont {Uwatoko}},\ }\bibfield  {title} {\bibinfo {title}
  {{Transport, Thermal, and Magnetic Properties of Heavy Fermion Compound
  Ce$_3$TiBi$_5$}},\ }\href {https://doi.org/10.7566/JPSCP.38.011084}
  {\bibfield  {journal} {\bibinfo  {journal} {JPS Conference Proceedings}\
  }\textbf {\bibinfo {volume} {38}},\ \bibinfo {pages} {011084} (\bibinfo
  {year} {2023})}\BibitemShut {NoStop}%
\bibitem [{\citenamefont {Kwasigroch}\ \emph {et~al.}(2022)\citenamefont
  {Kwasigroch}, \citenamefont {Hu}, \citenamefont {Kr{\"{u}}ger},\ and\
  \citenamefont {Green}}]{Kwasigroch2022}%
  \BibitemOpen
  \bibfield  {author} {\bibinfo {author} {\bibfnamefont {M.~P.}\ \bibnamefont
  {Kwasigroch}}, \bibinfo {author} {\bibfnamefont {H.}~\bibnamefont {Hu}},
  \bibinfo {author} {\bibfnamefont {F.}~\bibnamefont {Kr{\"{u}}ger}},\ and\
  \bibinfo {author} {\bibfnamefont {A.~G.}\ \bibnamefont {Green}},\ }\bibfield
  {title} {\bibinfo {title} {{Magnetic hard-direction ordering in anisotropic
  Kondo systems}},\ }\href {https://doi.org/10.1103/PhysRevB.105.224418}
  {\bibfield  {journal} {\bibinfo  {journal} {Physical Review B}\ }\textbf
  {\bibinfo {volume} {105}},\ \bibinfo {pages} {224418} (\bibinfo {year}
  {2022})}\BibitemShut {NoStop}%
\bibitem [{\citenamefont {Shinozaki}\ \emph {et~al.}(2020)\citenamefont
  {Shinozaki}, \citenamefont {Motoyama}, \citenamefont {Tsubouchi},
  \citenamefont {Sezaki}, \citenamefont {Gouchi}, \citenamefont {Nishigori},
  \citenamefont {Mutou}, \citenamefont {Yamaguchi}, \citenamefont {Fujiwara},
  \citenamefont {Miyoshi},\ and\ \citenamefont {Uwatoko}}]{Shinozaki2020}%
  \BibitemOpen
  \bibfield  {author} {\bibinfo {author} {\bibfnamefont {M.}~\bibnamefont
  {Shinozaki}}, \bibinfo {author} {\bibfnamefont {G.}~\bibnamefont {Motoyama}},
  \bibinfo {author} {\bibfnamefont {M.}~\bibnamefont {Tsubouchi}}, \bibinfo
  {author} {\bibfnamefont {M.}~\bibnamefont {Sezaki}}, \bibinfo {author}
  {\bibfnamefont {J.}~\bibnamefont {Gouchi}}, \bibinfo {author} {\bibfnamefont
  {S.}~\bibnamefont {Nishigori}}, \bibinfo {author} {\bibfnamefont
  {T.}~\bibnamefont {Mutou}}, \bibinfo {author} {\bibfnamefont
  {A.}~\bibnamefont {Yamaguchi}}, \bibinfo {author} {\bibfnamefont
  {K.}~\bibnamefont {Fujiwara}}, \bibinfo {author} {\bibfnamefont
  {K.}~\bibnamefont {Miyoshi}},\ and\ \bibinfo {author} {\bibfnamefont
  {Y.}~\bibnamefont {Uwatoko}},\ }\bibfield  {title} {\bibinfo {title}
  {{Magnetoelectric Effect in the Antiferromagnetic Ordered State of
  Ce$_3$TiBi$_5$ with Ce Zig-Zag Chains}},\ }\href
  {https://doi.org/10.7566/JPSJ.89.033703} {\bibfield  {journal} {\bibinfo
  {journal} {Journal of the Physical Society of Japan}\ }\textbf {\bibinfo
  {volume} {89}},\ \bibinfo {pages} {033703} (\bibinfo {year}
  {2020})}\BibitemShut {NoStop}%
\bibitem [{\citenamefont {Bollore}\ \emph {et~al.}(1995)\citenamefont
  {Bollore}, \citenamefont {Ferguson}, \citenamefont {Hushagen},\ and\
  \citenamefont {Mar}}]{Bollore1995}%
  \BibitemOpen
  \bibfield  {author} {\bibinfo {author} {\bibfnamefont {G.}~\bibnamefont
  {Bollore}}, \bibinfo {author} {\bibfnamefont {M.~J.}\ \bibnamefont
  {Ferguson}}, \bibinfo {author} {\bibfnamefont {R.~W.}\ \bibnamefont
  {Hushagen}},\ and\ \bibinfo {author} {\bibfnamefont {A.}~\bibnamefont
  {Mar}},\ }\bibfield  {title} {\bibinfo {title} {{New Ternary Rare-Earth
  Transition-Metal Antimonides $RE_3M$Sb$_5$ ($RE$ = La, Ce, Pr, Nd, Sm; $M$ =
  Ti, Zr, Hf, Nb)}},\ }\href {https://doi.org/10.1021/cm00060a005} {\bibfield
  {journal} {\bibinfo  {journal} {Chemistry of Materials}\ }\textbf {\bibinfo
  {volume} {7}},\ \bibinfo {pages} {2229} (\bibinfo {year} {1995})}\BibitemShut
  {NoStop}%
\bibitem [{\citenamefont {Moore}\ \emph {et~al.}(2002)\citenamefont {Moore},
  \citenamefont {Deakin}, \citenamefont {Ferguson},\ and\ \citenamefont
  {Mar}}]{Moore2002a}%
  \BibitemOpen
  \bibfield  {author} {\bibinfo {author} {\bibfnamefont {S.~H.~D.}\
  \bibnamefont {Moore}}, \bibinfo {author} {\bibfnamefont {L.}~\bibnamefont
  {Deakin}}, \bibinfo {author} {\bibfnamefont {M.~J.}\ \bibnamefont
  {Ferguson}},\ and\ \bibinfo {author} {\bibfnamefont {A.}~\bibnamefont
  {Mar}},\ }\bibfield  {title} {\bibinfo {title} {{Physical Properties and
  Bonding in $RE_3$TiSb$_5$ ($RE$ = La, Ce, Pr, Nd, Sm)}},\ }\href
  {https://doi.org/10.1021/cm020731t} {\bibfield  {journal} {\bibinfo
  {journal} {Chemistry of Materials}\ }\textbf {\bibinfo {volume} {14}},\
  \bibinfo {pages} {4867} (\bibinfo {year} {2002})}\BibitemShut {NoStop}%
\bibitem [{\citenamefont {Matin}\ \emph {et~al.}(2017)\citenamefont {Matin},
  \citenamefont {Kulkarni}, \citenamefont {Thamizhavel}, \citenamefont {Dhar},
  \citenamefont {Provino},\ and\ \citenamefont {Manfrinetti}}]{Matin2017}%
  \BibitemOpen
  \bibfield  {author} {\bibinfo {author} {\bibfnamefont {M.}~\bibnamefont
  {Matin}}, \bibinfo {author} {\bibfnamefont {R.}~\bibnamefont {Kulkarni}},
  \bibinfo {author} {\bibfnamefont {A.}~\bibnamefont {Thamizhavel}}, \bibinfo
  {author} {\bibfnamefont {S.~K.}\ \bibnamefont {Dhar}}, \bibinfo {author}
  {\bibfnamefont {A.}~\bibnamefont {Provino}},\ and\ \bibinfo {author}
  {\bibfnamefont {P.}~\bibnamefont {Manfrinetti}},\ }\bibfield  {title}
  {\bibinfo {title} {{Probing the magnetic ground state of single crystalline
  Ce$_3$TiSb$_5$}},\ }\href {https://doi.org/10.1088/1361-648X/aa57c0}
  {\bibfield  {journal} {\bibinfo  {journal} {Journal of Physics: Condensed
  Matter}\ }\textbf {\bibinfo {volume} {29}},\ \bibinfo {pages} {145601}
  (\bibinfo {year} {2017})}\BibitemShut {NoStop}%
\bibitem [{\citenamefont {Ritter}\ \emph {et~al.}(2021)\citenamefont {Ritter},
  \citenamefont {Pathak}, \citenamefont {Filippone}, \citenamefont {Provino},
  \citenamefont {Dhar},\ and\ \citenamefont {Manfrinetti}}]{Ritter2021}%
  \BibitemOpen
  \bibfield  {author} {\bibinfo {author} {\bibfnamefont {C.}~\bibnamefont
  {Ritter}}, \bibinfo {author} {\bibfnamefont {A.~K.}\ \bibnamefont {Pathak}},
  \bibinfo {author} {\bibfnamefont {R.}~\bibnamefont {Filippone}}, \bibinfo
  {author} {\bibfnamefont {A.}~\bibnamefont {Provino}}, \bibinfo {author}
  {\bibfnamefont {S.~K.}\ \bibnamefont {Dhar}},\ and\ \bibinfo {author}
  {\bibfnamefont {P.}~\bibnamefont {Manfrinetti}},\ }\bibfield  {title}
  {\bibinfo {title} {{Magnetic ground states of Ce$_3$TiSb$_5$, Pr$_3$TiSb$_5$
  and Nd$_3$TiSb$_5$ determined by neutron powder diffraction and magnetic
  measurements}},\ }\href {https://doi.org/10.1088/1361-648X/abe9db} {\bibfield
   {journal} {\bibinfo  {journal} {Journal of Physics: Condensed Matter}\
  }\textbf {\bibinfo {volume} {33}},\ \bibinfo {pages} {245801} (\bibinfo
  {year} {2021})}\BibitemShut {NoStop}%
\bibitem [{\citenamefont {Murakami}\ \emph {et~al.}(2017)\citenamefont
  {Murakami}, \citenamefont {Yamamoto}, \citenamefont {Takeiri}, \citenamefont
  {Nakano},\ and\ \citenamefont {Kageyama}}]{Murakami2017}%
  \BibitemOpen
  \bibfield  {author} {\bibinfo {author} {\bibfnamefont {T.}~\bibnamefont
  {Murakami}}, \bibinfo {author} {\bibfnamefont {T.}~\bibnamefont {Yamamoto}},
  \bibinfo {author} {\bibfnamefont {F.}~\bibnamefont {Takeiri}}, \bibinfo
  {author} {\bibfnamefont {K.}~\bibnamefont {Nakano}},\ and\ \bibinfo {author}
  {\bibfnamefont {H.}~\bibnamefont {Kageyama}},\ }\bibfield  {title} {\bibinfo
  {title} {{Hypervalent Bismuthides La$_3$MBi$_5$ (M = Ti, Zr, Hf) and Related
  Antimonides: Absence of Superconductivity}},\ }\href
  {https://doi.org/10.1021/acs.inorgchem.7b00031} {\bibfield  {journal}
  {\bibinfo  {journal} {Inorganic Chemistry}\ }\textbf {\bibinfo {volume}
  {56}},\ \bibinfo {pages} {5041} (\bibinfo {year} {2017})}\BibitemShut
  {NoStop}%
\bibitem [{\citenamefont {Motoyama}\ \emph {et~al.}(2020)\citenamefont
  {Motoyama}, \citenamefont {Shinozaki}, \citenamefont {Tsubouchi},
  \citenamefont {Kuninaka}, \citenamefont {Nishigori}, \citenamefont {Miyoshi},
  \citenamefont {Fujiwara}, \citenamefont {Gouchi},\ and\ \citenamefont
  {Uwatoko}}]{Motoyama2020}%
  \BibitemOpen
  \bibfield  {author} {\bibinfo {author} {\bibfnamefont {G.}~\bibnamefont
  {Motoyama}}, \bibinfo {author} {\bibfnamefont {M.}~\bibnamefont {Shinozaki}},
  \bibinfo {author} {\bibfnamefont {M.}~\bibnamefont {Tsubouchi}}, \bibinfo
  {author} {\bibfnamefont {M.}~\bibnamefont {Kuninaka}}, \bibinfo {author}
  {\bibfnamefont {S.}~\bibnamefont {Nishigori}}, \bibinfo {author}
  {\bibfnamefont {K.}~\bibnamefont {Miyoshi}}, \bibinfo {author} {\bibfnamefont
  {K.}~\bibnamefont {Fujiwara}}, \bibinfo {author} {\bibfnamefont
  {J.}~\bibnamefont {Gouchi}},\ and\ \bibinfo {author} {\bibfnamefont
  {Y.}~\bibnamefont {Uwatoko}},\ }\bibfield  {title} {\bibinfo {title}
  {{Magnetic Properties of New Antiferromagnetic Compound of Ce$_3$ZrBi$_5$}},\
  }\href {https://doi.org/10.7566/JPSCP.30.011180} {\bibfield  {journal}
  {\bibinfo  {journal} {JPS Conference Proceedings}\ }\textbf {\bibinfo
  {volume} {30}},\ \bibinfo {pages} {011180} (\bibinfo {year}
  {2020})}\BibitemShut {NoStop}%
\bibitem [{\citenamefont {Nakagawa}\ \emph {et~al.}(2023)\citenamefont
  {Nakagawa}, \citenamefont {Shinozaki}, \citenamefont {Motoyama},
  \citenamefont {Nishigori}, \citenamefont {Fujiwara}, \citenamefont {Manago},\
  and\ \citenamefont {Miyoshi}}]{Nakagawa2023}%
  \BibitemOpen
  \bibfield  {author} {\bibinfo {author} {\bibfnamefont {K.}~\bibnamefont
  {Nakagawa}}, \bibinfo {author} {\bibfnamefont {M.}~\bibnamefont {Shinozaki}},
  \bibinfo {author} {\bibfnamefont {G.}~\bibnamefont {Motoyama}}, \bibinfo
  {author} {\bibfnamefont {S.}~\bibnamefont {Nishigori}}, \bibinfo {author}
  {\bibfnamefont {K.}~\bibnamefont {Fujiwara}}, \bibinfo {author}
  {\bibfnamefont {M.}~\bibnamefont {Manago}},\ and\ \bibinfo {author}
  {\bibfnamefont {K.}~\bibnamefont {Miyoshi}},\ }\bibfield  {title} {\bibinfo
  {title} {{Single Crystal Growth of Ce$_3$ZrSb$_5$ and Characterization of the
  Physical Properties}},\ }\href {https://doi.org/10.7566/JPSCP.38.011083}
  {\bibfield  {journal} {\bibinfo  {journal} {JPS Conference Proceedings}\
  }\textbf {\bibinfo {volume} {38}},\ \bibinfo {pages} {011083} (\bibinfo
  {year} {2023})}\BibitemShut {NoStop}%
\bibitem [{\citenamefont {Hayami}\ and\ \citenamefont
  {Kusunose}(2022)}]{Hayami2022a}%
  \BibitemOpen
  \bibfield  {author} {\bibinfo {author} {\bibfnamefont {S.}~\bibnamefont
  {Hayami}}\ and\ \bibinfo {author} {\bibfnamefont {H.}~\bibnamefont
  {Kusunose}},\ }\bibfield  {title} {\bibinfo {title} {{Magnetic Toroidal
  Moment under Partial Magnetic Order in Hexagonal Zigzag-Chain Compound
  Ce$_3$TiBi$_5$}},\ }\href {https://doi.org/10.7566/JPSJ.91.123701} {\bibfield
   {journal} {\bibinfo  {journal} {Journal of the Physical Society of Japan}\
  }\textbf {\bibinfo {volume} {91}},\ \bibinfo {pages} {123701} (\bibinfo
  {year} {2022})}\BibitemShut {NoStop}%
\bibitem [{\citenamefont {Rodr{\'{i}}guez-Carvajal}(1993)}]{Fullprof}%
  \BibitemOpen
  \bibfield  {author} {\bibinfo {author} {\bibfnamefont {J.}~\bibnamefont
  {Rodr{\'{i}}guez-Carvajal}},\ }\bibfield  {title} {\bibinfo {title} {{Recent
  advances in magnetic structure determination by neutron powder
  diffraction}},\ }\href {https://doi.org/10.1016/0921-4526(93)90108-I}
  {\bibfield  {journal} {\bibinfo  {journal} {Physica B: Condensed Matter}\
  }\textbf {\bibinfo {volume} {192}},\ \bibinfo {pages} {55} (\bibinfo {year}
  {1993})}\BibitemShut {NoStop}%
\bibitem [{\citenamefont {Momma}\ and\ \citenamefont {Izumi}(2011)}]{VESTA}%
  \BibitemOpen
  \bibfield  {author} {\bibinfo {author} {\bibfnamefont {K.}~\bibnamefont
  {Momma}}\ and\ \bibinfo {author} {\bibfnamefont {F.}~\bibnamefont {Izumi}},\
  }\bibfield  {title} {\bibinfo {title} {{VESTA 3 for three-dimensional
  visualization of crystal, volumetric and morphology data}},\ }\href
  {https://doi.org/10.1107/S0021889811038970} {\bibfield  {journal} {\bibinfo
  {journal} {Journal of Applied Crystallography}\ }\textbf {\bibinfo {volume}
  {44}},\ \bibinfo {pages} {1272} (\bibinfo {year} {2011})}\BibitemShut
  {NoStop}%
\bibitem [{\citenamefont {Stokes}\ \emph {et~al.}()\citenamefont {Stokes},
  \citenamefont {Hatch},\ and\ \citenamefont {Campbell}}]{StokesISOTROPY}%
  \BibitemOpen
  \bibfield  {author} {\bibinfo {author} {\bibfnamefont {H.~T.}\ \bibnamefont
  {Stokes}}, \bibinfo {author} {\bibfnamefont {D.~M.}\ \bibnamefont {Hatch}},\
  and\ \bibinfo {author} {\bibfnamefont {B.~J.}\ \bibnamefont {Campbell}},\
  }\href {iso.byu.edu} {\bibinfo {title} {{ISODISTORT, ISOTROPY Software
  Suite}}}\BibitemShut {NoStop}%
\bibitem [{\citenamefont {Campbell}\ \emph {et~al.}(2006)\citenamefont
  {Campbell}, \citenamefont {Stokes}, \citenamefont {Tanner},\ and\
  \citenamefont {Hatch}}]{Campbell2006}%
  \BibitemOpen
  \bibfield  {author} {\bibinfo {author} {\bibfnamefont {B.~J.}\ \bibnamefont
  {Campbell}}, \bibinfo {author} {\bibfnamefont {H.~T.}\ \bibnamefont
  {Stokes}}, \bibinfo {author} {\bibfnamefont {D.~E.}\ \bibnamefont {Tanner}},\
  and\ \bibinfo {author} {\bibfnamefont {D.~M.}\ \bibnamefont {Hatch}},\
  }\bibfield  {title} {\bibinfo {title} {{ISODISPLACE: A web-based tool for
  exploring structural distortions}},\ }\href
  {https://doi.org/10.1107/S0021889806014075} {\bibfield  {journal} {\bibinfo
  {journal} {Journal of Applied Crystallography}\ }\textbf {\bibinfo {volume}
  {39}},\ \bibinfo {pages} {607} (\bibinfo {year} {2006})}\BibitemShut
  {NoStop}%
\bibitem [{\citenamefont {Perez-Mato}\ \emph {et~al.}(2012)\citenamefont
  {Perez-Mato}, \citenamefont {Ribeiro}, \citenamefont {Petricek},\ and\
  \citenamefont {Aroyo}}]{Perez-Mato2012}%
  \BibitemOpen
  \bibfield  {author} {\bibinfo {author} {\bibfnamefont {J.~M.}\ \bibnamefont
  {Perez-Mato}}, \bibinfo {author} {\bibfnamefont {J.~L.}\ \bibnamefont
  {Ribeiro}}, \bibinfo {author} {\bibfnamefont {V.}~\bibnamefont {Petricek}},\
  and\ \bibinfo {author} {\bibfnamefont {M.~I.}\ \bibnamefont {Aroyo}},\
  }\bibfield  {title} {\bibinfo {title} {{Magnetic superspace groups and
  symmetry constraints in incommensurate magnetic phases}},\ }\href
  {https://doi.org/10.1088/0953-8984/24/16/163201} {\bibfield  {journal}
  {\bibinfo  {journal} {Journal of Physics: Condensed Matter}\ }\textbf
  {\bibinfo {volume} {24}},\ \bibinfo {pages} {163201} (\bibinfo {year}
  {2012})}\BibitemShut {NoStop}%
\bibitem [{\citenamefont {Biffin}\ \emph {et~al.}(2014)\citenamefont {Biffin},
  \citenamefont {Johnson}, \citenamefont {Choi}, \citenamefont {Freund},
  \citenamefont {Manni}, \citenamefont {Bombardi}, \citenamefont {Manuel},
  \citenamefont {Gegenwart},\ and\ \citenamefont {Coldea}}]{Biffin2014}%
  \BibitemOpen
  \bibfield  {author} {\bibinfo {author} {\bibfnamefont {A.}~\bibnamefont
  {Biffin}}, \bibinfo {author} {\bibfnamefont {R.~D.}\ \bibnamefont {Johnson}},
  \bibinfo {author} {\bibfnamefont {S.}~\bibnamefont {Choi}}, \bibinfo {author}
  {\bibfnamefont {F.}~\bibnamefont {Freund}}, \bibinfo {author} {\bibfnamefont
  {S.}~\bibnamefont {Manni}}, \bibinfo {author} {\bibfnamefont
  {A.}~\bibnamefont {Bombardi}}, \bibinfo {author} {\bibfnamefont
  {P.}~\bibnamefont {Manuel}}, \bibinfo {author} {\bibfnamefont
  {P.}~\bibnamefont {Gegenwart}},\ and\ \bibinfo {author} {\bibfnamefont
  {R.}~\bibnamefont {Coldea}},\ }\bibfield  {title} {\bibinfo {title}
  {{Unconventional magnetic order on the hyperhoneycomb Kitaev lattice in
  $\beta$-Li$_2$IrO$_3$: Full solution via magnetic resonant x-ray
  diffraction}},\ }\href {https://doi.org/10.1103/PhysRevB.90.205116}
  {\bibfield  {journal} {\bibinfo  {journal} {Physical Review B}\ }\textbf
  {\bibinfo {volume} {90}},\ \bibinfo {pages} {205116} (\bibinfo {year}
  {2014})}\BibitemShut {NoStop}%
\bibitem [{\citenamefont {Hafner}\ \emph {et~al.}(2019)\citenamefont {Hafner},
  \citenamefont {Rai}, \citenamefont {Banda}, \citenamefont {Kliemt},
  \citenamefont {Krellner}, \citenamefont {Sichelschmidt}, \citenamefont
  {Morosan}, \citenamefont {Geibel},\ and\ \citenamefont
  {Brando}}]{Hafner2019}%
  \BibitemOpen
  \bibfield  {author} {\bibinfo {author} {\bibfnamefont {D.}~\bibnamefont
  {Hafner}}, \bibinfo {author} {\bibfnamefont {B.~K.}\ \bibnamefont {Rai}},
  \bibinfo {author} {\bibfnamefont {J.}~\bibnamefont {Banda}}, \bibinfo
  {author} {\bibfnamefont {K.}~\bibnamefont {Kliemt}}, \bibinfo {author}
  {\bibfnamefont {C.}~\bibnamefont {Krellner}}, \bibinfo {author}
  {\bibfnamefont {J.}~\bibnamefont {Sichelschmidt}}, \bibinfo {author}
  {\bibfnamefont {E.}~\bibnamefont {Morosan}}, \bibinfo {author} {\bibfnamefont
  {C.}~\bibnamefont {Geibel}},\ and\ \bibinfo {author} {\bibfnamefont
  {M.}~\bibnamefont {Brando}},\ }\bibfield  {title} {\bibinfo {title}
  {{Kondo-lattice ferromagnets and their peculiar order along the magnetically
  hard axis determined by the crystalline electric field}},\ }\href
  {https://doi.org/10.1103/PhysRevB.99.201109} {\bibfield  {journal} {\bibinfo
  {journal} {Physical Review B}\ }\textbf {\bibinfo {volume} {99}},\ \bibinfo
  {pages} {201109} (\bibinfo {year} {2019})}\BibitemShut {NoStop}%
\bibitem [{\citenamefont {Kr{\"{u}}ger}\ \emph {et~al.}(2014)\citenamefont
  {Kr{\"{u}}ger}, \citenamefont {Pedder},\ and\ \citenamefont
  {Green}}]{Kruger2014}%
  \BibitemOpen
  \bibfield  {author} {\bibinfo {author} {\bibfnamefont {F.}~\bibnamefont
  {Kr{\"{u}}ger}}, \bibinfo {author} {\bibfnamefont {C.~J.}\ \bibnamefont
  {Pedder}},\ and\ \bibinfo {author} {\bibfnamefont {A.~G.}\ \bibnamefont
  {Green}},\ }\bibfield  {title} {\bibinfo {title} {{Fluctuation-Driven
  Magnetic Hard-Axis Ordering in Metallic Ferromagnets}},\ }\href
  {https://doi.org/10.1103/PhysRevLett.113.147001} {\bibfield  {journal}
  {\bibinfo  {journal} {Physical Review Letters}\ }\textbf {\bibinfo {volume}
  {113}},\ \bibinfo {pages} {147001} (\bibinfo {year} {2014})}\BibitemShut
  {NoStop}%
\bibitem [{\citenamefont {Mazzone}\ \emph {et~al.}(2019)\citenamefont
  {Mazzone}, \citenamefont {Gauthier}, \citenamefont {Maimone}, \citenamefont
  {Yadav}, \citenamefont {Bartkowiak}, \citenamefont {Gavilano}, \citenamefont
  {Raymond}, \citenamefont {Pomjakushin}, \citenamefont {Casati}, \citenamefont
  {Revay}, \citenamefont {Lapertot}, \citenamefont {Sibille},\ and\
  \citenamefont {Kenzelmann}}]{Mazzone2019a}%
  \BibitemOpen
  \bibfield  {author} {\bibinfo {author} {\bibfnamefont {D.~G.}\ \bibnamefont
  {Mazzone}}, \bibinfo {author} {\bibfnamefont {N.}~\bibnamefont {Gauthier}},
  \bibinfo {author} {\bibfnamefont {D.~T.}\ \bibnamefont {Maimone}}, \bibinfo
  {author} {\bibfnamefont {R.}~\bibnamefont {Yadav}}, \bibinfo {author}
  {\bibfnamefont {M.}~\bibnamefont {Bartkowiak}}, \bibinfo {author}
  {\bibfnamefont {J.~L.}\ \bibnamefont {Gavilano}}, \bibinfo {author}
  {\bibfnamefont {S.}~\bibnamefont {Raymond}}, \bibinfo {author} {\bibfnamefont
  {V.}~\bibnamefont {Pomjakushin}}, \bibinfo {author} {\bibfnamefont
  {N.}~\bibnamefont {Casati}}, \bibinfo {author} {\bibfnamefont
  {Z.}~\bibnamefont {Revay}}, \bibinfo {author} {\bibfnamefont
  {G.}~\bibnamefont {Lapertot}}, \bibinfo {author} {\bibfnamefont
  {R.}~\bibnamefont {Sibille}},\ and\ \bibinfo {author} {\bibfnamefont
  {M.}~\bibnamefont {Kenzelmann}},\ }\bibfield  {title} {\bibinfo {title}
  {{Evolution of Magnetic Order from the Localized to the Itinerant Limit}},\
  }\href {https://doi.org/10.1103/PhysRevLett.123.097201} {\bibfield  {journal}
  {\bibinfo  {journal} {Physical Review Letters}\ }\textbf {\bibinfo {volume}
  {123}},\ \bibinfo {pages} {097201} (\bibinfo {year} {2019})}\BibitemShut
  {NoStop}%
\end{thebibliography}

\end{document}